\documentclass[aps,prb,floatfix,twocolumn,notitlepage,superscriptaddress,10pt]{revtex4-2}
\usepackage{xcolor}
\usepackage{grffile}
\usepackage{amsmath, amsthm, amssymb, bbold}
\usepackage[normalem]{ulem}
\usepackage{cancel}
\usepackage{bm}
\usepackage{microtype}
\usepackage{hyperref}
\usepackage{setspace}
\usepackage{grffile}
\hypersetup{linkcolor=blue,urlcolor=blue,citecolor=blue}
\usepackage{amsmath}    
\usepackage{amssymb}
\usepackage{graphicx}   
\usepackage{verbatim}   
\usepackage{color}      
\usepackage{microtype}
\usepackage[normalem]{ulem}
\usepackage{natbib}
\usepackage{enumitem} 
\usepackage{dsfont} 
\usepackage{hyperref}   

\newcommand{\ket}[1]{|#1\rangle}
\newcommand{\average}[1]{\langle #1\rangle}

\newcommand{\dbra}[1]{\langle\!\langle #1}
\newcommand{\dket}[1]{|#1\rangle\!\rangle}

\begin{document}

\title{Quantum Coherence in a Maximally Hot Hubbard Chain}

\author{C\u at\u alin Pa\c scu Moca}
\email{mocap@uoradea.ro}
\affiliation{Department of Physics, University of Oradea,  410087, Oradea, Romania}
\affiliation{Department of Theoretical Physics, Institute of Physics, Budapest University of Technology and Economics, M\H{u}egyetem rkp.~3, H-1111 Budapest, Hungary}
\author{Ovidiu I. P\^ a\c tu}
\affiliation{Institute for Space Sciences, Bucharest-M\u agurele, R 077125, Romania}
\author{Bal\'azs D\'ora}
\affiliation{Department of Theoretical Physics, Institute of Physics, Budapest University of Technology and Economics, M\H{u}egyetem rkp.~3, H-1111 Budapest, Hungary}
\affiliation{MTA-BME Lend\"ulet "Momentum" Open Quantum Systems Research Group, Institute of Physics, Budapest University of Technology and Economics,
M\H uegyetem rkp. 3., H-1111, Budapest, Hungary}
\author{Gergely Zar\'and}
\affiliation{Department of Theoretical Physics, Institute of Physics, Budapest University of Technology and Economics, M\H{u}egyetem rkp.~3, H-1111 Budapest, Hungary}

\date{\today}
\begin{abstract}

We present a detailed study of the real-time dynamics and spectral properties of the one-dimensional 
fermionic Hubbard model at infinite temperature. Using tensor network simulations in Liouville space, 
we compute the single-particle Green’s function and analyze its dynamics across a broad range of interaction 
strengths. To complement the time-domain approach, we develop a high-resolution Chebyshev expansion 
method within the density matrix formalism, enabling direct access to spectral functions in the frequency domain. 
In the non-interacting limit, we derive exact analytical expressions for the Green’s function, providing a benchmark 
for our numerical methods. As interactions are introduced, we observe a transition in the spectral function from a 
sharp peak at the free dispersion to a broadened two-band structure associated with hole and doublon excitations. 
These features are well captured by a Hubbard-I mean-field approximation, even at intermediate coupling. 
At infinite interaction strength ($U = \infty$), we exploit a determinant representation of the Green’s function 
to access both real-time and spectral properties. In this regime, the system retains a sharp, cosine-like 
momentum dispersion in frequency space, while the dynamics display nontrivial light-cone spreading 
with sub-ballistic scaling. Our results demonstrate that strong correlations and nontrivial 
quantum coherence can persist even at infinite temperature.
\end{abstract}
\maketitle

\section{Introduction}

Understanding the real-time dynamics and spectral properties of strongly correlated quantum systems 
remains one of the central challenges in modern condensed matter and quantum many-body physics~\cite{moeckel2009real,polkovnikov2011colloquium}. 
Among the wide variety of correlated systems, the Hubbard model stands as a paradigmatic example, 
capturing the essential competition between kinetic energy and interaction effects~\cite{georges1996dynamical,essler2005one,eckstein2010interaction,qin2022hubbard,arovas2022hubbard}.

Studies of the spectral functions in the Hubbard model~\cite{hubbard1963electron} have revealed distinct behaviors at low and high 
temperatures, reflecting the complex interplay between electron correlations and thermal fluctuations~\cite{dong2019dynamical,peters2015local,kadow2022hole}. At low temperatures, the spectral function exhibits features associated with Fermi liquid behavior 
in the weakly interacting regime~\cite{benfatto2006fermi} and with spin-charge separation in the strongly interacting, 
one-dimensional limit~\cite{voit1995one,giamarchi2003quantum,essler2005one,markhof2016spectral}. In particular, in one dimension, 
the emergence of separate spinon and holon branches in the spectral function has been confirmed using methods such as density 
matrix renormalization group (DMRG) and exact diagonalization~\cite{benthien2004spectral,benthien2007spin,carmelo2006general,
prelovvsek2013ground}. At half-filling and strong coupling, the system enters a Mott insulating phase characterized by the opening 
of a gap in the single-particle spectrum~\cite{essler2005one}. As temperature increases, thermal excitations smear out these 
low-energy features, leading to the suppression of coherent quasiparticle peaks and the appearance of broad incoherent 
structures~\cite{nocera2018finite,tindall2019heating,soltanieh2014spectral,rashid2023thermal,pactu2024exact}.

In this work, we investigate the single-particle spectral function of the one-dimensional Hubbard model in the 
infinite-temperature limit, examining its evolution from the non-interacting case to the regime of strong interactions. This 
infinite-temperature limit, despite its conceptual simplicity, remains largely unexplored in the context of spectral properties, 
offering new insights into the nature of incoherent excitations and the underlying dynamics. To address this problem, we employ a 
tensor network representation of the density matrix~\cite{kshetrimayum2019tensor,verstraete2008matrix,montangero2018introduction,
schollwock2011density,schollwock2013matrix}, relying on a vectorized (or “superfermion”) formalism~\cite{dzhioev2011super} where 
the Lindblad-like coherent evolution of the system is efficiently simulated using matrix product states (MPS)\cite{paeckel2019time}. This enables us to access the real-time dynamics of non-equilibrium Green’s functions even in the presence of 
strong interactions\cite{van2022momentum,tian2021matrix,chen2024matrix,white2008spectral}. In parallel, we explore an alternative 
frequency-domain approach based on Chebyshev polynomial expansions~\cite{braun2014numerical,halimeh2015chebyshev,wolf2015spectral,wolf2014chebyshev,weisse2006kernel,chen2023topological,zhao2023chebyshev,ganahl2014chebyshev,weisse2008chebyshev,
holzner2011chebyshev}, which bypasses the limitations imposed by finite-time Fourier transforms and improves spectral resolution. 
While the Chebyshev expansion method is well-established for computing spectral functions through 
resolvent techniques, its application to the direct evolution of the density matrix has not been explored previously. In this 
work, we introduce this novel extension within the density matrix formalism for studying spectral properties of interacting 
quantum systems at infinite temperatures.

We begin by examining the exactly solvable case of $U = 0$, where the Green’s function can be expressed in closed form using 
Bessel functions. This allows for precise benchmarking of both the real-time and Chebyshev expansion methods. We then turn to the 
interacting regime ($U > 0$), where the real-time Green’s function exhibits exponential decay within the light cone. Notably, the 
light cone’s boundary displays Kardar-Parisi-Zhang scaling (KPZ)~\cite{kardar1986dynamic,corwin2016kardar,ljubotina2019kardar}, $x \sim t^{2/3}$. 
In the frequency domain, interactions lead to a broadening 
and splitting of the spectral function, giving rise to an emergent two-band structure.

At the same time, the impenetrable limit of the Hubbard chain ($U=\infty$) provides a unique 
window into strongly correlated dynamics at infinite temperature. In this regime, double 
occupancies are projected out, rendering the model exactly solvable while still exhibiting highly 
nontrivial physics. We show that the single-particle Green’s function retains a sharp light-cone 
structure, yet the spreading front deviates from both purely diffusive ($x \propto t^{1/2}$) and 
KPZ-type ($x \propto t^{2/3}$) scaling, instead revealing an intermediate, super-diffusive 
behavior. Furthermore, the local Green’s function decays anomalously as a power law $G(x=0,t) \sim 
t^{-0.85}$, signaling a distinct dynamical universality class.

\section{Theoretical framework}
\subsection{Hubbard model}
The one-dimensional Hubbard model is a paradigmatic lattice model of interacting fermions, widely used to investigate the interplay of kinetic energy and electron-electron interactions in strongly correlated systems~\cite{hubbard1963electron,essler2005one}. It captures essential features of Mott insulating behavior, spin-charge separation, and magnetic ordering, and serves as a minimal model for understanding correlated metals, insulators, or even high-temperature superconductors.
The model is defined by the Hamiltonian
\begin{gather}
	H = -{J\over 2} \sum_{x=-L/2}^{L/2} \sum_{\sigma = \uparrow, \downarrow} \left( c^\dagger_{x\sigma} c_{x+1,\sigma} + \text{h.c.} \right) \nonumber\\
	+ U \sum_{x=-L/2}^{L/2} \left(n_{x\uparrow}-{1\over 2}\right)\left(n_{x\downarrow}-{1\over 2}\right),
	\label{eq:Hamiltonian}
\end{gather}
where $J$ is the nearest-neighbor hopping amplitude, and $U$ denotes the strength of the local repulsive interaction between fermions of opposite spin on the same site. The operators $c^\dagger_{x\sigma}$  create a fermion with spin $\sigma$ at site $x$, while $n_{x\sigma} = c^\dagger_{x\sigma} c_{x\sigma}$ is the corresponding number operator.

One of the main objectives of this work is to compute Green's functions in both real and momentum space, as functions of time or frequency, 
and to construct their corresponding spectral representations~\cite{seabra2014real,tian2021matrix}. To achieve this, we employ 
multiple computational strategies, all of which using the fact that at infinite temperature the density matrix is trivial, and
proportional with the identity matrix, $\rho \propto \mathbb{1}$. 

We focus on the evaluation of the time- and position-dependent lesser component of the Green's functions involving local fermionic or bosonic operators $\mathcal{O}_{x\sigma}$,
\begin{gather}
 G_{\sigma}^{<}	(x,t) = (-i)^s \langle \mathcal{O^\dagger}_{0\sigma}(0) \mathcal{O}_{x\sigma}(t) \rangle \nonumber \\
 \phantom{aaaaaaaaaa}= (-i)^s\, \text{Tr} \{ \mathcal{O^\dagger}_{0\sigma}(0) \mathcal{O}_{x\sigma}(t) \},
 \label{eq:G_x_t}
\end{gather}
where $s = \pm 1$ for bosonic (fermionic) operators, respectively. 
Assuming translational invariance, we perform a spatial Fourier transform to 
obtain the momentum- and time-resolved Green's function
\begin{gather}
	G^{<}_\sigma(k,t) = \frac{1}{L} \sum_{x} e^{ikx} G^{<}_\sigma(x,t),
	\label{eq:G_k_t}
\end{gather}
where the momentum $k$ is quantized as $k = 2\pi n / L$ with $n = -L/2, \dots, L/2$, and the resolution is $\Delta k = 2\pi / L$ due to the finite lattice size.
To explore spectral properties, we further Fourier transform in time to obtain the momentum- and frequency-dependent Green's function
\begin{gather}
	G_\sigma^{<}(k,\omega) = \frac{T_t}{N_t} \sum_t e^{i\omega t} G_\sigma^{<}(k,t),
\end{gather}
where $T_t$ is the maximum simulation time and $N_t$ is the number of time points sampled. The frequencies are quantized as $\omega = 2\pi n / T_t$ with $n = -T_t/2, \dots, T_t/2$, yielding a frequency resolution of $\Delta \omega = 2\pi / T_t$.

We similarly evaluate the greater component of the Green’s function, defined as
\begin{gather}
G_{\sigma}^{>}(x,t) = -i \langle \mathcal{O}_{x\sigma}(t) \mathcal{O}^\dagger_{0\sigma}(0) \rangle,
\label{eq:G_x_t>}
\end{gather}
which, together with the lesser component, enables the construction of the retarded Green’s function via
\begin{gather}
G_{\sigma}^{R}(x,t) = \theta(t)\left[ G^>_{\sigma}(x,t) - G^<_{\sigma}(x,t) \right].
\label{eq:G_x_t>ret}
\end{gather}
From the Fourier transform of $G^R_{\sigma}(x,t)$, we obtain the spectral function
\begin{equation}
A_{\sigma}(k,\omega) = -\frac{1}{\pi} \, \text{Im}\, G^{R}_{\sigma}(k,\omega),
\label{eq:A_k_w}
\end{equation}
which characterizes the momentum-and frequency-resolved single-particle excitation 
spectrum. This quantity provides insight into how electronic excitations propagate 
through the system and how interactions reshape the underlying band structure.

We primarily focus on the single-particle Green’s functions, where the central operators 
are $\mathcal{O}_{x\sigma}(t) = c_{x\sigma}(t)$. 

The momentum-integrated single-particle spectral function,  it is defined as
\begin{equation}
A_\sigma(\omega) = \frac{2\pi}{L} \sum_k A_\sigma(k,\omega),
\label{eq:A_int}
\end{equation}
This quantity is particularly useful for examining the overall spectral weight distribution 
and identifying broad features such as the formation of Hubbard bands. Unlike the momentum-resolved
 spectral function $A(k,\omega)$, which reveals dispersion relations and coherence properties of 
 quasiparticles, the integrated function emphasizes the energy-resolved structure and is typically
  more robust against finite-size effects or symmetry-breaking perturbations.
The momentum-integrated spectral function $A_\sigma(\omega)$ 
is otherwise identical to the local density of states (LDOS), $\rho_\sigma(\omega)$, defined as
\begin{gather}
\rho_\sigma(\omega) = -\frac{1}{\pi} \operatorname{Im} G_\sigma(x=0,\omega).
\end{gather}
Thus, both quantities capture the same local spectral information.

\begin{figure}[tbh!]
	\begin{center}
	 \includegraphics[width=0.9\columnwidth]{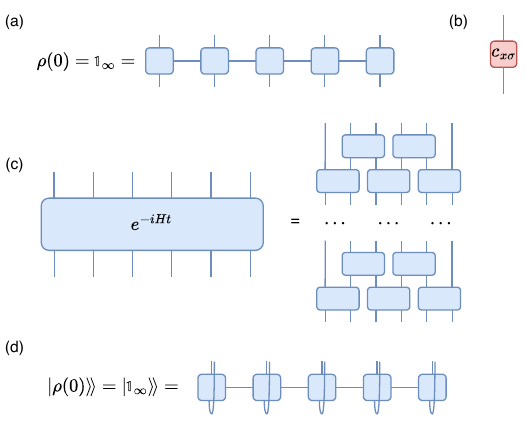}
	 \caption{Schematic representation of the tensor network structures used. (a) Matrix product operator (MPO) form of the density matrix. (b) Representation of a local operator acting at site $x$. (c) Trotter decomposition of the time-evolution operator into alternating layers of two-site gates acting on even and odd bonds.(d) The MPS representation of the vectorized density matrix.}
	 \label{fig:Dictionary}
	\end{center}
\end{figure}
\subsection{Operator evolution}\label{sec:OE}

To compute correlation functions efficiently, we rely on tensor network 
representations that allow for scalable simulations of quantum many-body systems. Fig.~\ref{fig:Dictionary} provides a schematic overview of the tensor network structures used in computing the Green’s functions. The density matrix, a central object in our approach, is represented as a Matrix Product Operator (MPO), as depicted in Fig.~\ref{fig:Dictionary}(a),
\begin{gather}
\rho = \sum_{\{\alpha_j,i_j,i'_j\}} 
    \mathrm{Tr}\left[ M^{i_{-{L/2}} i'_{-{L/2}}}_{\alpha_{-{L/ 2}}} 
                     M^{i_{-{L/ 2}+1} i'_{-{L/ 2}+1}}_{\alpha_{-{L/ 2}} \alpha_{-{L/ 2}+1}} \cdots
                     M^{\,i_{{L/ 2}} i'_{L/ 2}}_{\alpha_{{L/ 2}}} \right] \times\nonumber\\
    |i_{-L/2} i_{-L/2+1} \cdots i_{L/2}\rangle \langle i'_{-L/2} i'_{-L/2+1} \cdots i'_{L/2}|.
	\label{eq:rho_MPO}
\end{gather}
Here, each tensor $M^{i_\ell i'_\ell}$ encodes the local structure of $\rho$, with physical indices $(i_\ell, i'_\ell)$ and virtual (bond) indices $\alpha_{\ell-1}, \alpha_\ell$ connecting neighboring sites. At infinite temperature, the density matrix reduces to a completely mixed state
\begin{eqnarray}
\rho &=& \bigotimes_{j=-{L/2}}^{L/2} \mathbb{1}_j \\
	 &=& \sum_{\{i_j\}} |i_{-L/2} i_{-L/2+1} \cdots i_{L/2}\rangle \langle i_{-L/2} i_{-L/2+1} \cdots i_{L/2}|,\nonumber
\end{eqnarray}
which corresponds to an MPO with bond dimension 1, where each local tensor is simply the identity $M^{\,i_j i'_j} = \delta_{i_j i'_j}$.

\begin{figure}[tbh!]
	\begin{center}
	 \includegraphics[width=0.65\columnwidth]{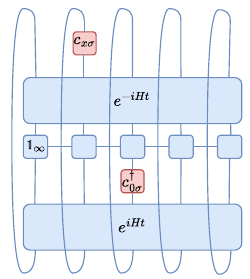}
	 \caption{ Tensor network representation used to compute the Green's function defined in Eq.~\eqref{eq:G_x_t}, involving two fermionic operators acting at sites $x$ and $0$. Although the annihilation (creation) operators are depicted as local, each is accompanied by a non-local Jordan-Wigner string implicit in the fermionic representation.}
	 \label{fig:Operator_evolution}
	\end{center}
\end{figure}

In the operator evolution approach, the Green's functions $G_\sigma^{\lessgtr}(x,t)$ are computed via the 
Heisenberg time evolution of the operators. The time evolution of $\mathcal{O}_{x\sigma}(t)$ is carried out 
using the Heisenberg picture for the operator evolution method~\cite{nocera2018finite}.
For example, the evaluation of the single particle Green's function, as defined  in Eq.~\eqref{eq:G_x_t>}, 
with fermionic 
operators $c_{x\sigma}$ and $c^\dagger_{0\sigma}$ acting at sites $x$ and $0$, 
respectively, is shown schematically in Fig.~\ref{fig:Operator_evolution}. Since we are 
interested in computing correlations for all positions $x$ along the chain, it is 
computationally advantageous to evolve only one operator—in particular, we evolve 
$c^\dagger_{0\sigma}$ backward in time. This strategy is illustrated in Fig.~\ref
{fig:Operator_evolution}.

The Green's function is then obtained by contracting the MPO for the operator $c_{x\sigma}$ with the time-evolved operator $c^\dagger_{0\sigma}(-t)$ and taking the trace
\begin{gather}
G^{>}_\sigma(x,t) = -i\,\text{Tr}\left\{ \left(e^{iHt} c_{x\sigma}e^{-iHt} \right)  c^\dagger_{0\sigma} \right\} \nonumber \\
\phantom{aaaaaaaa}= -i\,\text{Tr}\left\{ c_{x\sigma} \left( e^{-iHt} c^{\dagger}_{0\sigma} e^{iHt} \right) \right\}.
\end{gather}
This formulation allows us to compute the Green's function at all spatial separations 
$x$ from a single time evolution of the operator $c^\dagger_{0\sigma}$, making the computation 
efficient. This simplification is a direct consequence of the trivial structure 
of the density matrix at infinite temperature. The time evolution is performed using 
the time evolving block decimation (TEBD) 
approach~\cite{vidal2007classical} as implemented in the ITensor library~\cite{fishman2022itensor}. 

Importantly, the infinite-temperature setting isolates the contribution of the operator dynamics, decoupled 
from state-specific correlations. As a result, the method remains numerically efficient even for systems with 
strong interactions or large entanglement, offering a powerful tool for probing dynamical properties through 
operator evolution alone.

\subsection{Vectorized von-Neumann evolution}\label{sec:VvNE}
An alternative and equally powerful strategy for computing time-dependent Green's functions at infinite temperature is based on the vectorization of the density matrix~\cite{cui2015variational,jaschke2018one,casagrande2021analysis,nakano2021tensor}. In this approach, operator expectation values are evaluated by mapping the density operator to a pure state in a doubled Hilbert space, effectively transforming the problem into a standard state evolution calculation.
The infinite-temperature density matrix, $\rho \sim \mathbb{1}$, is vectorized as
\begin{equation}
\dket{\rho} = \sum_{\{i_j\}} |i_{-L/2} i_{-L/2+1} \cdots i_{L/2}\rangle \otimes |i_{-L/2} i_{-L/2+1} \cdots i_{L/2}\rangle,
\end{equation}
which corresponds to an MPS of bond dimension 1 with local tensors $\sim \delta_{i_j, i'_j}$.
This state is illustrated in Fig.~\ref{fig:Dictionary}(d). 

In this doubled formalism, any operator $\mathcal{O}$ acting on the system corresponds to a superoperator acting on the vectorized state
\begin{equation}
\text{Tr}\{\mathcal{O} \rho\} = \dbra{\rho} |\mathcal{O} \otimes \mathbb{1} \dket{\rho}.
\end{equation}
Time evolution under a Hamiltonian $H$ translates into a vectorized von-Neumann evolution
\begin{equation}
\dket{\rho(t)}= e^{-i(H \otimes \mathbb{1} - \mathbb{1} \otimes H)t} \dket{\rho}.
\end{equation}
This allows us to compute the Green's function from a standard MPS inner product
\begin{equation}
G^>_\sigma(x,t) = -i \dbra{\rho}| (c_{x\sigma} \otimes \mathbb{1}) 
e^{-i(H \otimes \mathbb{1} - \mathbb{1} \otimes H)t} 
(c^\dagger_{0\sigma} \otimes \mathbb{1}) \dket {\rho}.
\label{eq:G_x_t_vec}
\end{equation}
\begin{figure}[t!]
	\begin{center}
	 \includegraphics[width=0.65\columnwidth]{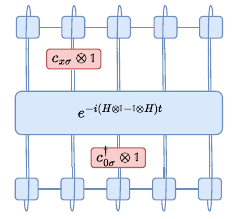}
	 \caption{  
	 Tensor network representation for computing the Green's function using the vectorized von-Neumann evolution according to Eq.~\eqref{eq:G_x_t_vec}. The density matrix is encoded as a matrix product state (MPS) in the doubled Hilbert space and time-evolved under the vectorized formalism.}
	 \label{fig:Vectorized_evolution}
	\end{center}
\end{figure}

This form is natural for the MPS simulations where the time evolution is performed using Trotterized two-site gates on the doubled system, as illustrated in Fig.~\ref
{fig:Vectorized_evolution}. Importantly, due to the infinite-temperature assumption, the 
vectorized state $\dket{\rho}$ has no initial entanglement build in, which keeps the computational cost low.
This construction, while conceptually straightforward, does not permit proper 
use of Abelian or non-Abelian symmetries, 
since the infinite-temperature state lacks well-defined quantum numbers~\cite{moca2022simulating}. 
To address this limitation, it is advantageous to introduce a dual set of fermionic 
operators, $\tilde{c}_{x\sigma}$, which act on a dual Fock space~\cite{dzhioev2011super, dzhioev2012nonequilibrium}. 
These operators obey standard anticommutation relations, $\{ \tilde{c}_{x\sigma}, \tilde{c}^\dagger_{x'\sigma'} \} = \delta_{x,x'}\delta_{\sigma,\sigma'}$. Within this extended framework, often referred to as the \emph{superfermion representation}, the density matrix evolves according to  
\begin{equation}
	\dket{\rho(t)} = e^{-i (H - \tilde{H})t} \dket{\rho},
\end{equation}
where $\tilde{H}$ is the same Hamiltonian as $H$—e.g., Eq.~\eqref{eq:Hamiltonian}—but expressed 
in terms of the tilde (dual) operators. For a single site, the Hilbert space consists of 16 states, 
associated with 
\begin{equation}
	\ket {\alpha, \tilde \alpha } = \{\ket{0}, \ket{\uparrow},\ket{\downarrow}, 
	\ket{\uparrow\downarrow} \}\otimes \{\ket{\tilde 0}, \ket{\tilde \uparrow},\ket{\tilde \downarrow}, \ket{\tilde \uparrow\tilde \downarrow} \} 
\end{equation}
In this construction, the first index is associate with the regular space, while the second 
	index is associated with the dual(tilde) space. 
This formalism can be interpreted as a basis transformation in the doubled Fock space, where 
both the physical and dual (tilde) degrees of freedom are explicitly included. In this 
representation, the infinite-temperature density matrix takes a particularly simple and 
factorized form  
\begin{eqnarray}
	\dket{\rho} & = &\bigotimes_{j=-L/2}^{L/2} \exp\left\{-i\sum_{\sigma} c^\dagger_{j\sigma} \tilde{c}^\dagger_{j\sigma} \right\} \ket{0,\tilde{0}} \nonumber \\
	&=& \bigotimes_{j=-L/2}^{L/2} \dket{\rho_j}, \label{eq:GS}
\end{eqnarray}
where $\ket{0,\tilde{0}}$ is the vacuum of the doubled Fock space. Each local contribution 
$\dket{\rho_j}$ to the full vectorized state can be written explicitly as
\begin{gather}
	\dket{\rho_j} = \frac{1}{2}\left( \ket{0,\tilde{0}}_j - i\ket{\uparrow,\tilde{\uparrow}}_j 
	- i\ket{\downarrow,\tilde{\downarrow}}_j + \ket{\uparrow\downarrow,\tilde{\uparrow}\tilde{\downarrow}}_j \right).
	\label{eq:rho_j}
\end{gather}
Constructed in this way, the vectorized density matrix $\dket{\rho}$ plays the role of a 
"reference" or "ground" state at infinite temperature, with respect to which expectation values 
are computed in the standard inner product. For instance, one finds $\langle n_{j\sigma} \rangle = 1/2$, $\langle \tilde{n}_{j\sigma} \rangle = 1/2$, and the total occupation defined as ${\cal N}_{j\sigma} = n_{j\sigma} - \tilde{n}_{j\sigma}$ satisfies $\langle {\cal N}_{j\sigma} \rangle = 0$. The minus sign in the definition of ${\cal N}$ ensures that the total conserved charge vanishes for the infinite-temperature state, i.e., ${\cal N} = 0$ for $\dket{\rho}$, 
and each of the states forming $\dket{\rho_j}$ in Eq.~\eqref{eq:rho_j} has zero charge~\cite{moca2022simulating}.

\subsection{Chebyshev polynomial approach}
The two approaches discussed above are particularly well-suited for computing the Green's function in 
real space and real time. When using open boundary conditions, however, the accessible simulation time 
is constrained by the emergence of a light cone: once quasiparticles reach the system edges, boundary 
reflections introduce interference effects that degrade the quality of the data. As a result, the 
frequency resolution is fundamentally limited by the maximum evolution time $T$, with a  frequency 
resolution of $\Delta \omega = 2\pi/T$.

To overcome this limitation, we can compute the spectral function directly in momentum and frequency 
by using an extension of the Chebyshev polynomial approach constructed to work with the density matrix. This approach borrows from the vectorized von-Neumann evolution methods. 
For that, we adopt 
the Chebyshev polynomial expansion, a method that allows one to avoid explicit time 
evolution and instead approximate spectral quantities through recursive polynomial constructions. 
The imaginary part of the Green's function can be represented in Lehmann form and expanded using 
Chebyshev polynomials $T_n({\cal L})$ of a rescaled Liouvillian operator without the dephasing part 
${\cal L} \propto H \otimes \mathbb{1} - \mathbb{1} \otimes H$ with spectrum in $[-1, 1]$ ~\cite{holzner2011chebyshev,ganahl2014chebyshev}. 
To carry out the rescaling procedure, we begin by estimating the spectral 
extremities $L_{\text{min}}$ and $L_{\text{max}}$ of the operator $\mathcal{L}$ 
through two independent DMRG calculations. These bounds are then used to define a linear rescaling transformation with parameters
\begin{equation}
a = \frac{L_{\text{max}} - L_{\text{min}}}{2 - \delta}, \qquad b = \frac{L_{\text{max}} + L_{\text{min}}}{2},
\end{equation}
where a small offset $\delta \sim 0.01$ is introduced to ensure that the rescaled spectrum lies 
strictly within the interval $[-1,1]$. This constraint is important for the numerical stability of 
the Chebyshev polynomial recursion. The same linear transformation is applied to the physical 
frequencies via $\omega = a\, \omega’ + b$, with the rescaled frequency variable $\omega’$ confined 
to the interval $[-1,1]$.
This expansion becomes
\begin{equation}
\text{Im}\, G^>_\sigma(k,\omega) \approx {1\over a} \frac{1}{\sqrt{1 - \omega'^2}} \left[ \mu_0 + 2 \sum_{n=1}^{N_{\text{max}}} \mu_n T_n(\omega') \right],
\end{equation}
with the Chebyshev moments
\begin{equation}
\mu_n = \dbra{\rho} | (c_{k\sigma} \otimes \mathbb{1}) T_n({\cal L}) (c^\dagger_{k\sigma} \otimes \mathbb{1}) \dket{\rho}.
\label{eq:mu_n}
\end{equation}
These moments are efficiently computed using recursive relations of the Chebyshev polynomials
\begin{align}
\dket{b_0} &= (c_{k\sigma} \otimes \mathbb{1}) \dket{\rho}, \nonumber \\
|b_1\rangle &= {\cal L} |b_0\rangle, \\
\dket{b_{n+1}} &= 2 {\cal L}\dket{b_n} - \dket{b_{n-1}},\nonumber
\end{align}
with $\mu_n = \dbra{ b_0} \dket{b_n}$.
As the state $\dket{\rho}$ is a product state, the action of the operator $c_{k\sigma}$  in Eq.~\eqref{eq:mu_n} can be efficiently carried out within the MPS/MPO framework using standard 
tensor network techniques~\cite{fishman2022itensor}. Moreover, the absence of entanglement in the 
initial state and the nature of the Chebyshev recursion ensure 
that the bond dimensions remain controlled, even in the presence of interactions, offering 
significantly enhanced resolution and 
accuracy in the frequency domain~\cite{wolf2014chebyshev, wolf2015spectral}.  In this work, we 
further enhance the method by integrating the superfermion representation into the framework, 
allowing for a more efficient treatment within the density matrix formalism.
To improve convergence and mitigate Gibbs oscillations due to finite expansion 
order $N_{\text{max}}$, spectral broadening can be 
introduced using smoothing kernels such as the Jackson kernel~\cite{silver1996kernel,
weisse2006kernel}, leading to well-behaved 
approximations of the spectral function with tunable resolution. This makes the Chebyshev method 
especially suited for computing $A(k,\omega)$, across a wide range of frequencies with high 
precision, even in interacting systems where time evolution becomes 
challenging. However, because the calculation resolves momentum explicitly, an independent 
Chebyshev recursion must be carried out for each momentum value $k$. 

\section{Spectral and Dynamical Properties}

\subsection{Single particle spectral functions}
\begin{figure*}[tbh]
	\begin{center}
	 \includegraphics[width=1.95\columnwidth]{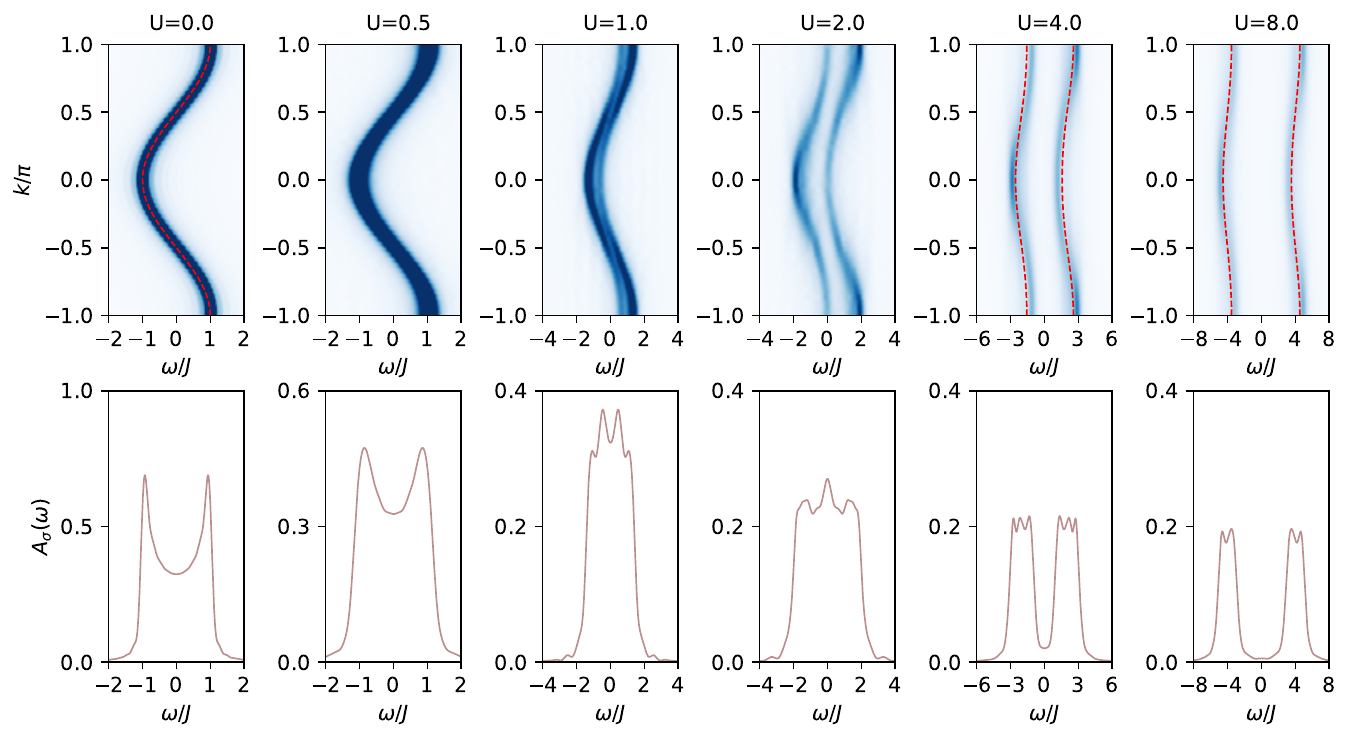}
	 \caption{
Top row: Momentum-resolved single-particle spectral function $A_\sigma(k,\omega)$ for different interaction strengths $U$, as indicated in each panel. The data are obtained using the Chebyshev expansion method on a chain of length $L = 50$ and using $M=2000$ moments. 
For $U = 0$, the dashed red lines indicate the exact non-interacting dispersion, given by $\varepsilon_k = -J \cos(k)$. In the cases of $U = 4,8$, 
the dashed lines correspond to the mean-field dispersion relation, capturing the interaction-induced band renormalization in the strong coupling regime. 
Bottom row: Corresponding momentum-integrated spectral functions $A_\sigma(\omega)$, as defined in Eq.~\eqref{eq:A_int}. For $U = 0$, 
the spectrum forms a single, continuous band spanning $\omega \in [-J, J]$, with sharp resonances near the band edges arising 
from contributions near momenta $k=\pm \pi$ and $k=0$. As $U$ increases, the spectral weight redistributes and begins to 
separate. In the strong-coupling limit ($U \gg J$), the spectrum clearly splits into two distinct bands: a lower band 
dominated by holon-like excitations, an upper band associated with doublon-like excitations.}
	 \label{fig:A_k_w}
	\end{center}
\end{figure*}

In the non-interacting limit, $U = 0$, the single-particle spectral function can be computed 
analytically due to the quadratic form of the Hubbard Hamiltonian. As a result, the single-particle Green’s 
function can be evaluated exactly  and the spectral 
function takes on a well-known form,
\begin{equation}
A^{(0)}_\sigma(k,\omega) = \delta(\omega - \varepsilon_k).
\label{eq:A0_k_w}
\end{equation}
This expression represents a sharp delta function peak at the single-particle energy $\varepsilon_k = 
-J\cos(k)$, characteristic of a system with well-defined quasiparticles. It provides a valuable 
benchmark for validating numerical simulations and serves as a reference against which 
interaction-driven features at finite $U$ can be identified.

As shown in the corresponding panel of Fig.~\ref{fig:A_k_w}, the numerically computed single-particle 
spectral function $A^{(0)}_\sigma(k,\omega)$ using the Chebyshev expansion method accurately 
reproduces the expected delta-function structure. The spectral weight is strictly confined to the 
energy range $\omega \in [-J, J]$, consistent with the non-interacting band structure. The sharp 
resonances at the band-edges are generated by the contributions 
near the momenta $k=\pm \pi$ and $k=0$. 

As the interaction strength $U$ increases, the single-particle spectral function $A_\sigma(k,\omega)$ 
undergoes a qualitative transformation, the spectral weight begins to redistribute, and a gap opens 
between two emerging bands. These correspond to holon-like (lower Hubbard band) and doublon-like 
(upper Hubbard band) excitations. The separation between the bands grows with $U$, reflecting the 
energy cost of double occupancy. In the strong-coupling regime ($U \gg J$), the spectral function 
clearly splits into two well-defined Hubbard bands, centered approximately at $\omega \approx \pm U/2$, with their internal dispersions determined by the kinetic energy scale $J$.

Important features of the single-particle spectral function $A_\sigma(k,\omega)$ can be understood 
using a mean-field level analysis, reminiscent of the Hubbard-I approximation~\cite
{hubbard1963electron,dorneich2000strong,nocera2018finite}. A key insight comes from decomposing the 
fermionic annihilation operator $c_{i\sigma}$ into contributions associated with distinct local 
occupancy sectors, effectively separating processes involving doublons (sites occupied by both spins) 
and holons (empty opposite-spin sites). This decomposition reads
\begin{equation}
c_{x\sigma} = c_{x\sigma} n_{x\bar\sigma} \tilde{n}_{x\bar\sigma} \tilde{n}_{x\sigma} + c_{x\sigma} (1 - n_{x\bar\sigma})(1 - \tilde{n}_{x\bar\sigma}) \tilde{n}_{x\sigma},
\label{eq:decomposition}
\end{equation}
where the first term corresponds to doublon-like configurations and the second to holon-like ones. Importantly, this splitting respects charge conservation in the vectorized Liouville-space formalism, where each term maintains the total particle number on the doubled Hilbert space.

Within this framework, the single-particle spectral function develops a two-band structure at 
finite and large $U$, reflecting the separation of low-energy hole-like excitations and 
high-energy doublon-like excitations. These features are well captured by the 
Hubbard-I approximation, which in the large $U$ limit predicts the following dispersions for the two bands
\begin{equation}
E_{\pm}(k) = \frac{1}{2} \left( \varepsilon_k \pm \sqrt{\varepsilon_k^2 + U^2} \right)\, ,
\label{eq:Ek}
\end{equation}
along with their corresponding momentum-dependent spectral weights:
\begin{equation}
Z_{\pm}(k) = \frac{1}{2} \left(1 \pm \frac{\varepsilon_k}{\sqrt{\varepsilon_k^2 + U^2}} \right).
\label{eq:Zk}
\end{equation}
These weights determine how the spectral intensity is distributed between the two bands: at $k = \pm \pi/2$, both bands carry equal weight, while away from this point, the balance shifts in 
favor of either the hole or doublon branch, depending on the sign of $\varepsilon_k$. The 
Hubbard-I approximation thus offers a qualitatively accurate picture of the spectral function 
even at moderate interactions ($U \sim 4$), capturing both the opening of a correlation-induced 
gap and the redistribution of spectral weight. As $U$ increases further, the spectrum clearly 
splits into upper and lower Hubbard bands, with increasing separation and sharper features, 
consistent with a strongly correlated regime. These predictions are confirmed by our numerically 
computed spectral functions, as shown in Fig.~\ref{fig:A_k_w}, where the overall band structure 
and weight distribution closely follow the mean-field expectations.

\subsection{Local density of states}
\begin{figure}[tbh!]
	\begin{center}
	 \includegraphics[width=0.95\columnwidth]{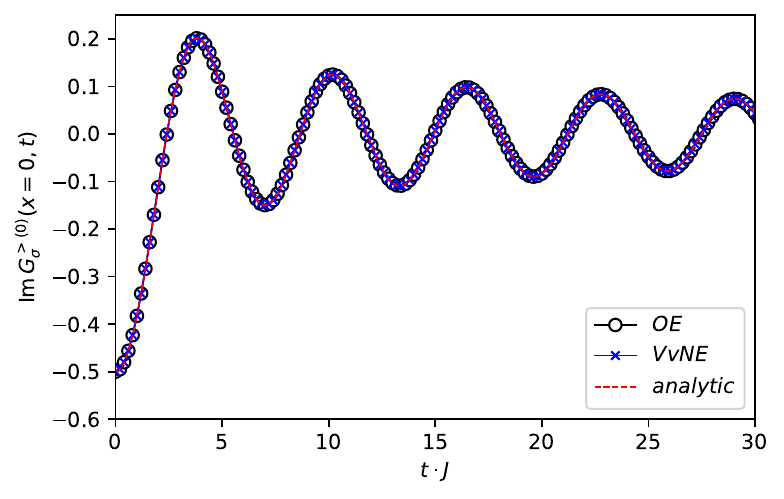}
	 \caption{Time evolution of the Green's function $G^{(0)}(x=0,t)$ in the middle of the chain 
	  The dashed red lines correspond to the analytical expression given in Eq.~\eqref{eq:G0_x_t}, while the symbols correspond to the operator evolution (OE) or vectorized von Neumann evolution (VvNE).}
	 \label{fig:G0_x_t}
	\end{center}
\end{figure}
We  computed the local  density of states (LDOS)  defined as 
which captures the local spectral weight at a given site and frequency. 
Figure~\ref{fig:A_k_w} (bottom row) presents results for $\rho_\sigma(\omega)$ using the Chebyshev expansion. For $U=0$ it reproduce  the analytical result obtained by Fourier transforming the time-domain expression in Eq.~\eqref{eq:G0_x_t}
\begin{gather}\label{eq:rho_0}
 A_\sigma(\omega) = \frac{1}{\pi}\frac{1}{\sqrt{J^2 - \omega^2}} \theta(J - |\omega|),
\end{gather}
demonstrating the characteristic square-root singularities at the band edges, $\omega = \pm J$, of the non-interacting density of states.

In the Chebyshev method, we employed $2000$ moments and applied Jackson kernel damping~\cite{silver1996kernel,
weisse2006kernel} to suppress Gibbs oscillations. 
As $U$ increases, these singularities are gradually smoothed out, and a two-peak  (Hubbard-band) structure emerges. The 
spectral weight splits into two bands centered  
approximately at $\pm U/2$, signaling the local suppression of charge fluctuations and the  
onset of strong correlations. These bands correspond to incoherent holon and doublon  
excitations propagating in a highly thermal background.

\subsection{Real-time dynamics}
\begin{figure*}[tbh!]
	\begin{center}
	 \includegraphics[width=1.9\columnwidth]{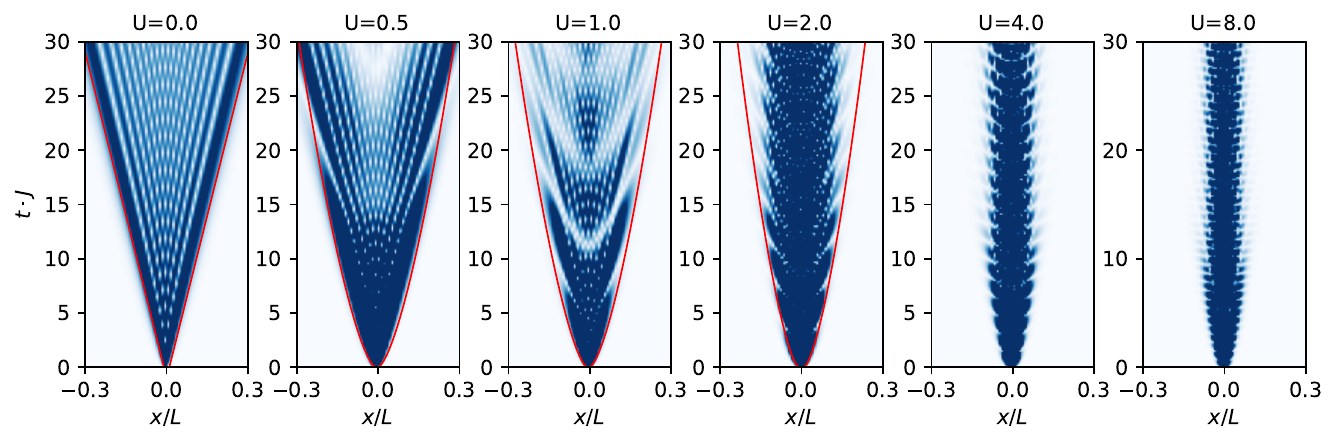}
	 \caption{
	Absolute value of the single particle Green’s function, $|G^{>}(x,t)|$, shown for various interaction strengths $U$ as labeled in each panel. 
	 At $U = 0$, a ballistic spreading of quasiparticles is clearly visible with $v=J$, 
	 indicated by the red line. For $U>0$ the edge of the lightcone shows 
	 a Kardar-Parisi-Zhang expansion as $x\propto t^{2/3}$. The results were obtained using the operator evolution approach. 
}
	 \label{fig:G_x_t}
	\end{center}
\end{figure*}
To analyze the time evolution of single-particle excitations, it is convenient to work in momentum space, where the non-interacting part of the Hubbard Hamiltonian in Eq.~\eqref{eq:Hamiltonian} becomes diagonal. The annihilation operator in real space is related to its momentum-space counterpart via a discrete Fourier transform
\begin{equation}
c_{x\sigma} = \frac{1}{\sqrt{L}} \sum_{k} e^{ikx} c_{k\sigma},
\label{eq:annihilation_op}
\end{equation}
with momenta $k = \frac{2\pi}{L} n$ for $n \in {-L/2, \dotsc, L/2 - 1}$ and periodic boundary conditions. In this basis, the kinetic Hamiltonian reads
\begin{equation}
H_0 = \sum_{k, \sigma} \varepsilon_k, c^\dagger_{k\sigma} c_{k\sigma}, \quad \text{where } \varepsilon_k = -J\cos(k),
\end{equation}
and the time evolution of the annihilation operator follows simply as
\begin{equation}
c_{k\sigma}(t) = c_{k\sigma}e^{-i\varepsilon_k t} = c_{k\sigma}e^{iJt\cos(k)}.
\end{equation}
Using Eq.~\eqref{eq:annihilation_op} along with standard properties of Bessel functions, the time evolution of the annihilation operator in real space takes the form
\begin{equation}
	c_{x'\sigma}(t) = \sum_{x} i^{x - x'} J_{x - x'}(\Omega t)\, c_{x\sigma},
\end{equation}
where $J_n(x)$ denotes the Bessel function of the first kind of order $n$, and the characteristic frequency is given by $\Omega = J$. 
At infinite temperature, the density matrix is proportional to the identity matrix, $\rho = \mathbb{1}/L$, and the expectation values are given by $\langle c^\dagger_{x\sigma} c_{x'\sigma'} \rangle_0 = \langle c_{x\sigma} c^\dagger_{x'\sigma'} \rangle_0 = {1\over 2}\delta_{xx'} \delta_{\sigma\sigma'}$, where, in the non-interacting limit the average is defined as $\langle \dots \rangle_0 = \text{Tr} \{ \rho \dots \}$. This allows us to provide an analytical expression for the time dependent, non-interacting Green's function, 
\begin{gather}
	G^{>(0)}_{\sigma}(x-x',t) = -i\average{c_{x\sigma}(t)c^\dagger_{x'\sigma'}(0)}_0\nonumber\\
	 \phantom{aaaaaaaaa}= -{1\over 2}i^{x'-x+1}J_{x'-x}(\Omega t)\delta_{\sigma,\sigma'},
	 \label{eq:G0_x_t}
\end{gather}
Note that correlations between sites separated by an odd number of lattice spacings are real, whereas those separated by an even number are purely imaginary. The superscript $(0)$ referes to the non-interacting Green's function. 

Building on the analytical expression for the time-dependent correlation function in Eq.~\eqref{eq:G0_x_t}, we can also characterize its long-time behavior by examining the asymptotics of the Bessel functions. For large arguments, the Bessel function of the first kind exhibits an oscillatory decay given by~\cite{abramowitz1965handbook}
\begin{equation}
	J_n(\Omega t) \sim \sqrt{\frac{2}{\pi \Omega t}} \cos\left(\Omega t - \frac{n\pi}{2} - \frac{\pi}{4}\right).
\end{equation}
Substituting this into Eq.~\eqref{eq:G0_x_t}, we find that the Green's function at infinite temperature decays as 
\begin{equation}
	G^{>(0)}_{\sigma}(x,t) \sim \frac{1}{\sqrt{\Omega t}} \cos\left(\Omega t - \frac{x\pi}{2} - \frac{\pi}{4}\right),
\end{equation}
modulated by an overall phase and a factor depending on the parity of the separation. 
This expression indicates that correlations decay algebraically with time, as 
\begin{equation}
G^{>(0)}_{\sigma}(x,t)\propto t^{-1/2}\,,\quad t\to \infty
\end{equation}
while retaining oscillatory features due to coherent propagation of quasiparticles. 


These analytical results provide a reference point for verifying the accuracy of the numerical 
techniques used in this work. For instance, Fig.\ref{fig:G0_x_t} presents a direct comparison 
between the analytical expression in Eq.\eqref{eq:G0_x_t} and the outcomes of our numerical 
simulations, demonstrating excellent agreement and thereby validating the correctness of our 
implementation.

\begin{figure}[tbh!]
	\begin{center}
	 \includegraphics[width=0.95\columnwidth]{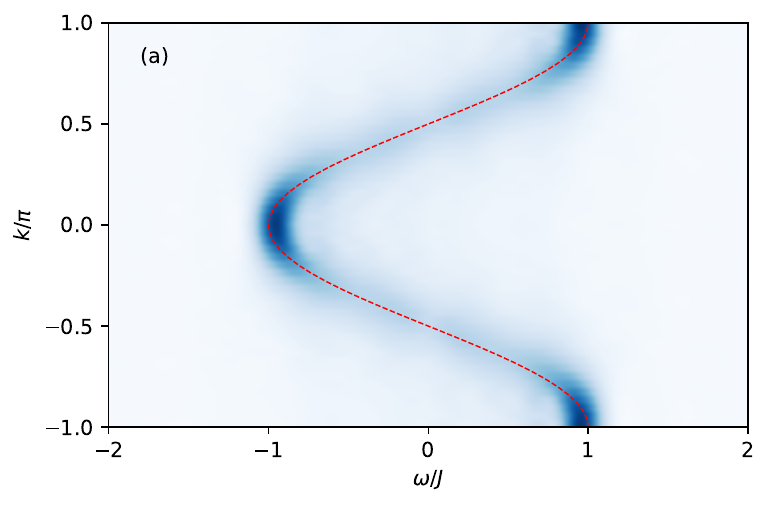}
	 \includegraphics[width=0.95\columnwidth]{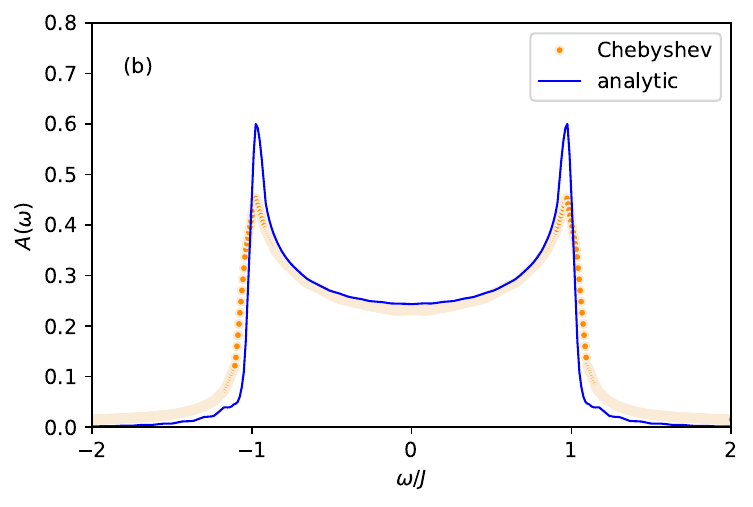}
	 \caption{
(a) Momentum-resolved single-particle spectral function $A_\sigma(k,\omega)$ in the $U = \infty$ limit. 
The red dashed curve indicates the non-interacting dispersion relation, $\varepsilon_k = -J\cos(k)$.
 Results were obtained using the Chebyshev expansion method with $M = 2000$ moments on a chain of length $L = 50$.
(b) Momentum-integrated spectral function $A_\sigma(\omega)$ at fixed filling $n = 2/3$. 
The symbols denote numerical data, while the solid line corresponds to the exact analytical 
result computed for a chain of length $L = 48$.}
	\label{fig:A_x_w_U_inf}
	\end{center}
\end{figure}
In Fig.~\ref{fig:G_x_t}, we present a density plot of the Green's function obtained using the 
operator evolution approach.

In the non-interacting limit, $U=0$, it clearly illustrates a ballistic spreading of non-interacting 
quasiparticles with velocity $v = J$, as reflected by the emergence of a well-defined light-cone 
structure.

In contrast to the non-interacting case ($U = 0$), where correlations spread ballistically, 
the edges of the light cone in the interacting Hubbard model follow Kardar-Parisi-Zhang 
(KPZ) scaling~\cite{ljubotina2019kardar,krajnik2020kardar}. Specifically, the front of the 
spreading correlations expands as $x \propto t^{2/3}$, a hallmark of KPZ dynamics, which is 
expected to emerge at infinite temperature~\cite{moca2023kardar}. This sub-ballistic 
scaling reveals that even at infinite temperature, the system retains nontrivial dynamical 
structure, governed by slow, yet coherent correlation spreading characterized by universal 
KPZ exponents.

\section{The $U=\infty$ limit}
In this section, we explore the single-particle spectral function in the analytically 
tractable limit of infinite on-site repulsion, $U = \infty$, at infinite temperature $T = \infty$. In this regime, the Hubbard model simplifies due to the complete 
suppression of doubly occupied states, which are excluded from the dynamics owing to their 
divergent energy cost~\cite{essler2005one,IPA98,pactu2024exact}. This impenetrable limit renders the model 
exactly solvable and thus provides a valuable reference point for benchmarking our tensor 
network-based numerical techniques.
In the $U \to \infty$ limit, double occupancies are energetically forbidden, and the 
system’s dynamics are confined to the subspace of states with at most single occupancy per 
site. The effective Hamiltonian governing the dynamics in this constrained Hilbert space is 
given by the projected form
\begin{gather}\label{eq:H_U_inf}
H^{\infty} = \mathcal{P} \left[ -\frac{J}{2} \sum_{x=-L/2}^{L/2-1} \sum_{\sigma = {\uparrow, \downarrow}} \left( c_{x\sigma}^\dagger c_{x+1\sigma} + \text{h.c.} \right) \right] \mathcal{P} , 
\end{gather}
where $\mathcal{P}$ is the Gutzwiller projector that enforces the no-double-occupancy constraint
\begin{gather}
\mathcal{P} = \prod_{x=-L/2}^{L/2} \left(1 - n_{x\uparrow} n_{x\downarrow} \right)  , \qquad \mathcal{P}^\dagger = \mathcal{P} ,  \qquad \mathcal{P}^2 = \mathcal{P}  .
\end{gather}
In this  limit,  the  wavefunctions of the Hubbard model 
factorize into spin and  charge components \cite{OgataShiba90,essler2005one}, allowing for the 
analytical derivation of determinant representations for the correlation functions. Details of the 
calculation are presented in Appendix~\ref{sec:Appendix}.

Figure~\ref{fig:A_x_w_U_inf} presents the momentum-resolved 
single-particle spectral function $A_\sigma(k,\omega)$ 
computed numerically via the Chebyshev expansion method. 
To assess the accuracy of our numerical approach, we also compare 
the momentum-integrated spectral function $A_\sigma(\omega)$ with the exact 
analytical result at the same filling and similar system size.
The agreement between the two approaches demonstrates the reliability and precision of 
the Chebyshev method in this strongly correlated, exactly solvable regime.
\begin{figure}[tbh!]
	\begin{center}
	 \includegraphics[width=0.95\columnwidth]{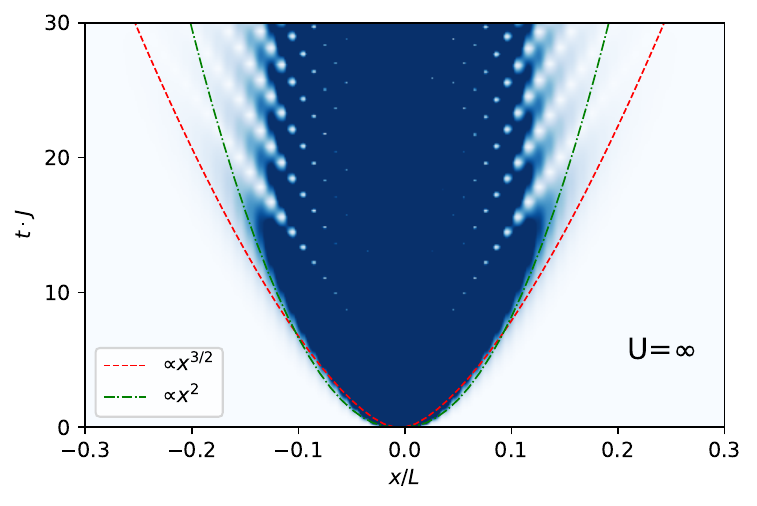}
	 \caption{Absolute value of the single particle Green’s function, $|G_\sigma^{>}(x,t)|$, 
	 for $U=\infty$. The edge of the light cone does not follow a perfect KPZ 
	 ($x \propto t^{2/3}$) or diffusive ($x \propto t^{1/2}$) scaling, but instead exhibits 
	 intermediate behavior. The results were obtained using the operator evolution approach.}
	  \label{fig:G_x_t_U_inf}
	\end{center}
\end{figure}
\begin{figure}[tbh!]
	\begin{center}
	 \includegraphics[width=0.95\columnwidth]{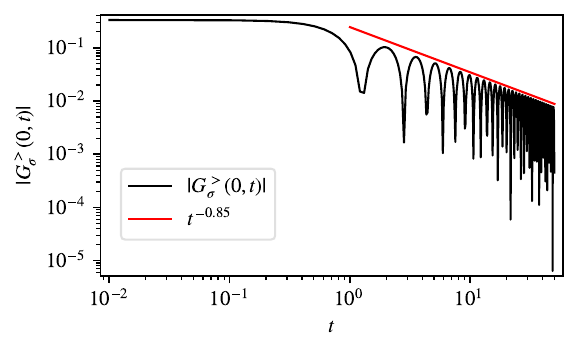}
	 \caption{
Absolute value of the single-particle Green’s function, $|G_\sigma^{>}(x=0,t)|$, evaluated at the 
center of the chain for $U=\infty$. The data reveal an anomalous power-law decay, $|G_\sigma^{>}
(x=0,t)| \propto t^{-0.85}$.}
	  \label{fig:G_0_t_U_inf}
	\end{center}
\end{figure}
In Fig.~\ref{fig:G_x_t_U_inf}, we show the real-space and time-dependent Green’s 
function $G_\sigma^{>}(x, t)$. It exhibits a clear light-cone structure, 
but the spreading front does not follow a perfect KPZ scaling ($x \propto t^{2/3}$) or
a purely diffusive behavior ($x \propto t^{1/2}$). 
This intermediate scaling indicates that the correlations propagate in a sub-ballistic yet 
super-diffusive fashion, pointing to an emergent dynamical regime where charge motion is 
constrained by the exclusion of doublons while spin fluctuations remain active. The resulting 
dynamics reflect a delicate balance between restricted fermionic transport and the fluctuating 
spin background, giving rise to correlation spreading that is anomalous compared to both 
free-fermion and diffusive expectations.

Complementary information is provided by the local Green’s function, $G_\sigma^{>}(x=0,t)$, which 
we find to decay as a nontrivial power law $|G_\sigma^{>}(x=0,t)| \propto t^{-0.85}$. This 
anomalous decay exponent deviates from the $t^{-0.5}$ behavior expected for free fermions and 
highlights the impact of strong correlations in the impenetrable limit. The  exponent 
signals enhanced persistence of local correlations, consistent with the constrained Hilbert space 
dynamics and the super-diffusive light-cone broadening. Together, the light-cone analysis and the 
local decay establish that the $U=\infty$ Hubbard model at infinite temperature exhibits genuinely 
anomalous dynamical behavior, distinct from both conventional diffusion and KPZ scaling.

\section{Conclusion}

We have analyzed the spectral properties of the one-dimensional Hubbard model 
at infinite temperature, focusing on the behavior of the single-particle Green’s 
function and its corresponding spectral function in both real time and frequency domains. 
We employed a combination of exact analytical results, time-evolution techniques in the Heisenberg picture, 
and frequency-resolved calculations based on the Chebyshev polynomial expansion. 
These complementary methods allowed us to validate our approach in the non-interacting 
limit and to explore the impact of interactions on the high-temperature spectral features of the model.

In the $U = 0$ case, we derived exact expressions for the Green’s function in 
terms of Bessel functions, which capture ballistic quasiparticle propagation 
and lead to a sharp spectral function peaked along the free-particle dispersion.
 Upon introducing interactions, the real-time dynamics become markedly different: 
 the spreading of correlations no longer follows the ballistic light cone but instead 
 exhibits a KPZ-type front with sub-ballistic scaling, consistent with recent 
 predictions for the high-temperature Hubbard chain.

In the frequency domain, we observed the formation of a two-band structure in the 
single-particle spectral function at finite interaction strengths, signaling the 
emergence of Hubbard bands even at infinite temperature. These bands are well described 
by the Hubbard-I approximation, which captures both the correct energy separation and 
spectral weights of the holon and doublon features. This agreement supports the notion 
that certain aspects of the zero-temperature Mott physics survive even in the maximally mixed thermal state.

The Chebyshev expansion technique proved particularly effective for computing 
local spectral functions, delivering excellent frequency resolution and 
capturing singular structures such as the square-root singularities at the band edges. 
Its numerical efficiency and precision make it a valuable tool for studying local 
dynamical quantities in strongly correlated systems, especially in the high-temperature regime where other methods may struggle.

In the impenetrable limit $U=\infty$, where the Hubbard model becomes exactly solvable, we 
uncovered an anomalous dynamical behavior at infinite temperature. The real-time Green’s function 
exhibits a light-cone structure with super-diffusive spreading, interpolating between diffusive 
and KPZ scaling, while the local correlations decay as a nontrivial power law $|G_\sigma^{>}(x=0,
t)| \propto t^{-0.85}$. These results demonstrate that, the system retains highly nontrivial 
dynamical signatures of strong correlations. The persistence of anomalous transport and slow local 
decay underscores that the $U=\infty$ Hubbard chain belongs to a distinct dynamical universality 
class, markedly different from both free-fermion and conventional diffusive systems.

Our findings highlight that infinite-temperature correlation functions retain 
rich signatures of interaction-driven physics typically associated with ground-state phenomena. 

\begin{acknowledgments}
This work received financial support from CNCS/CCCDI-UEFISCDI, under 
projects number PN-IV-P1-PCE-2023-0159 and PN-IV-P1-PCE-2023-0987 and by the National Research, Development and Innovation Office - NKFIH  Project Nos. K134437 and K142179.
We acknowledge the Digital Government Development and Project
Management Ltd.~for awarding us access to the Komondor HPC facility based in Hungary.
O.I.P. acknowledges financial support from Grant No. 30N/2023, provided through the National Core Program of the 
Romanian Ministry of Research, Innovation, and Digitization.

\end{acknowledgments}

\appendix
\section{The $U=\infty$, $T=\infty$ limit}\label{sec:Appendix}

The eigenstates of the Hamiltonian (\ref{eq:H_U_inf}) in the $(N,M)$-sector 
comprising $N$ particles of which $M$ have spin down, take the form
\begin{align}\label{app:eigen}
|\boldsymbol{\psi}_{N,M}(\boldsymbol{k},\boldsymbol{\lambda})\rangle&=\sum_{z_1,\cdots,z_N}
\sum_{\alpha_1,\cdots,\alpha_N=\{\uparrow,\downarrow\}}^{[N,M]}\,
\chi^{\boldsymbol{\alpha}}(z_1,\cdots,z_N|\boldsymbol{k},\boldsymbol{\lambda})\nonumber\\
& \qquad\qquad\qquad\times c_{z_N,\alpha_N}^\dagger \cdots c_{z_1,\alpha_1}^\dagger|0\rangle\, ,
\end{align}
where the notation $[N,M]$ over the summation sign indicates that the sum  is restricted to all 
configurations of $N$ spins with exactly $M$ spins down and $N-M$ spins up. Here 
$\boldsymbol{\alpha}=(\alpha_1,\cdots, \alpha_N)$ and  $|0\rangle$ denotes the Fock vacuum defined 
by  $c_{z,\alpha} |0\rangle=\langle 0|c_{z,\alpha}^\dagger=0$ for all values of $z$ and $\alpha$ 
and normalized such that $\langle 0|0\rangle=1$. The eigenstates in (\ref{app:eigen}) are 
parametrized  by two sets of distinct numbers: $\boldsymbol{k}=\{k_j\}_{j=1}^N$ which describes the 
charge degrees of freedom,  and $\boldsymbol{\lambda}=\{\lambda_j\}_{j=1}^M$, which describe the 
spin degrees of freedom. The wavefunction is given by  

\begin{figure}[tbh!]
	\begin{center}
	 \includegraphics[width=0.9\columnwidth]{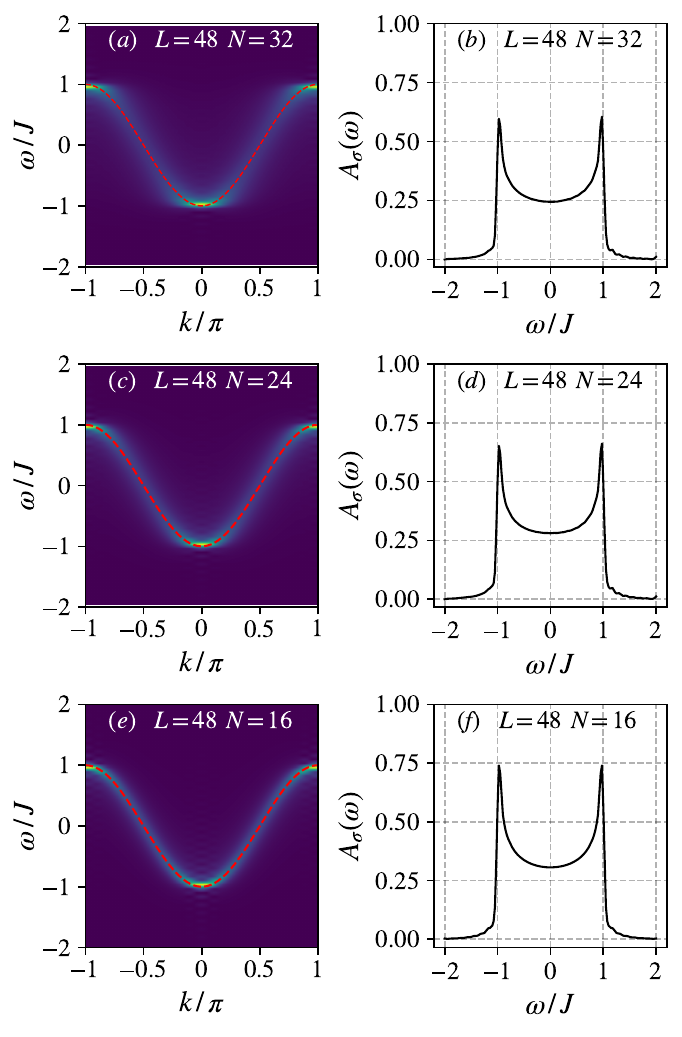}
	 \caption{(a,c,e) Momentum-resolved single-particle spectral function $A_\sigma(k,\omega)$ in the $U = \infty$ limit, 
               computed using the determinant formalism for various filling fractions indicated in each panel. 
               The dashed red lines indicate the exact non-interacting dispersion given by $\varepsilon(k)=-J \cos(k).$
               (b,d,f) Corresponding momentum-integrated spectral functions $A_\sigma(\omega)$ at the same fillings. }
	  \label{fig:A_k_w_exact}
	\end{center}
\end{figure}

\begin{align}\label{app:wavef}
\chi^{\boldsymbol{\alpha}}(\boldsymbol{z}|\boldsymbol{k},\boldsymbol{\lambda})=&\frac{1}{N!}
\left[\sum_{P\in S_N} \xi_{XX}^{(\boldsymbol{\alpha}, P\boldsymbol{\alpha})} (\boldsymbol{\lambda})
\theta(z_{P(1)}<\cdots <z_{P(N)})\right]\nonumber\\
&\qquad\qquad\qquad\qquad\times\det_N\left[\phi_{k_a}(z_b)\right]\, ,
\end{align}
where  $\theta(z_{P(1)}<\cdots <z_{P(N)})$  is a generalized Heaviside function which is equal to 
$1$ when $z_{P(1)}<\cdots <z_{P(N)}$ and $0$ otherwise. For a system with open boundary conditions 
on a lattice with $L$ sites the Slater determinant on the right hand side of Eq.~(\ref{app:wavef}) 
is built from the single particle orbitals  
\begin{equation}\label{app:spo}
\phi_k(z)=\left(\frac{2}{L+1}\right)^{1/2}\sin\left[\frac{\pi k z}{L+1}\right]\, ,\ \ k=1,\cdots,L.
\end{equation}
In the presence of an external potential, the structure of Eq.~(\ref{app:wavef}) remains unchanged, 
with the only difference being that the Slater determinant is constructed from the eigenfunctions of 
the corresponding single-particle Hamiltonian subject to the external potential. 
The spin sector is encoded in  $\xi_{XX}^{(\boldsymbol{\alpha}, P\boldsymbol{\alpha})} 
(\boldsymbol{\lambda})$, which are the wavefunctions of the $XX$ spin chain with periodic boundary 
conditions. They are given by \cite{Colomo1993}:
\begin{equation}\label{app:wavefxx}
\xi_{XX}^{(\boldsymbol{\alpha}, P\boldsymbol{\alpha})} (\boldsymbol{\lambda})=
\frac{1}{N^{M/2}}\left[\prod_{1\le j<l\le M} \mbox{sign}(n_l-n_j)\det_M\left(e^{i \lambda_a n_b}
\right)\right]\,\nonumber ,
\end{equation}
where $n_l$ denotes  the position of the $l$-th spin-down particle on the auxiliary lattice of the 
XX spin chain. The spin rapidities $\lambda_j$ satisfy the Bethe ansatz equations 
\begin{align}\label{app:baexxs}
e^{i\lambda_l N}&=(-1)^{M+1}\, ,\ \ l=1,\cdots,M\, .\
\end{align}
In the $U\rightarrow\infty$ limit the dynamics in the spin sector is frozen. As a result the 
time-dependent wavefunctions take the same form  as in (\ref{app:wavef}), 
with $\phi_k(z)$ replaced by $\phi_k(z,t)=e^{-i \varepsilon(k) t}\phi_k(z)$, where 
$\varepsilon(k)=-J\cos(k)$.

Even though the wavefunctions remain rather involved, one can derive determinant expressions for the 
correlation functions at arbitrary temperature using form-factor summation techniques 
\cite{KorepinSlavnov90,IP98}. Such determinant representations were obtained in 
\cite{IPA98, pactu2024exact} for the correlation functions
\begin{align}
g_\sigma^{(-)}(x,t;y,t')&=\langle c_{x\sigma}^\dagger(t)c_{y\sigma}(t')\rangle\, ,\\
g_\sigma^{(+)}(x,t;y,t')&=\langle c_{x\sigma}(t) c_{y\sigma}^\dagger(t')\rangle\, ,
\end{align}
which are related to the lesser and greater Green's functions, Eq.~(\ref{eq:G_x_t}) and 
(\ref{eq:G_x_t>}), via $G_\sigma^{<}(x,t)=-ig_{\sigma}^{(-)}(x,t;0,0)$ and 
$G_\sigma^{>}(x,t)=-ig_{\sigma}^{(+)}(x,t;0,0)\, .$ However, taking the infinite-temperature limit of 
the results derived in \cite{IPA98,pactu2024exact} yields correlation functions 
corresponding to a filling fraction $n=2/3$  due to the use of the grand-canonical ensemble. To obtain 
results valid for arbitrary filling fractions, one must instead compute the correlators within the 
canonical ensemble. Such canonical expressions are presented below.
For a system with $N$ fermions on a 
lattice with $L$ sites, the correlation functions have the following representations:
\begin{widetext}
\begin{align}
g_{\sigma}^{(-)}(x,t;y,t')&=\int_{0}^{2\pi}  d\phi\, \frac{e^{- i N\phi}}{2\pi}\,\int_{-\pi}^{\pi} \frac{F(\eta)}{2\pi}
\left[\det_L\left(\mathbf{1}+\frac{2}{3}\frac{e^{i\phi}}{Z^{1/N}} U^{(-)}(\eta)+\frac{1}{3}\frac{e^{i\phi}}{Z^{1/N}}R^{(-)} \right)\right.\nonumber\\
&\qquad\qquad\qquad\qquad\qquad\qquad\qquad\qquad\qquad\qquad\qquad  \left.-\det_L\left(\mathbf{1}+\frac{2}{3}\frac{e^{i\phi}}{Z^{1/N}} U^{(-)}(\eta) \right)\right]\, d\eta\, ,\label{app:detgmc}\\
g_{\sigma}^{(+)}(x,t;y,t')&=\int_{0}^{2\pi}  d\phi\, \frac{e^{- i N\phi}}{2\pi}\,\int_{-\pi}^{\pi} \frac{F(\eta)}{2\pi}
\left[\det_L\left(\mathbf{1}+\frac{2}{3}\frac{e^{i\phi}}{Z^{1/N}} U^{(+)}(\eta)-\frac{2}{3}\frac{e^{i\phi}}{Z^{1/N}} R^{(+)}(\eta) \right)\right.\nonumber\\
&\qquad\qquad\qquad\qquad\qquad\qquad\qquad\qquad\qquad\qquad\qquad  \left.+(g-1)\det_L\left(\mathbf{1}+\frac{2}{3}\frac{e^{i\phi}}{Z^{1/N}} U^{(+)}(\eta) \right)\right]\, d\eta\, ,\label{app:detgpc}
\end{align}
\end{widetext}
where $F(\eta)=3/(5-4\cos\eta)$, $Z=C^L_N2^N$ and 
\begin{equation}
g\equiv g(x,t;y,t')=\sum_{k=1}^L \phi_k(x,t)\phi^*_k(y,t')\, .
\end{equation}
The elements of the $L\times L$ matrices $U^{(\pm)}$ and $R^{(\pm)}$ are
defined in terms of the function
\begin{equation}
u(k,q|\eta,x,t)=\delta_{k,q}-\left[1-e^{- i \eta}\right]\sum_{z=x}^L \phi_k(z,t)\phi^*_q(z,t)\, ,
\end{equation}
and  given explicitly as follows (for $a,b=1,\cdots, L $):
\begin{align}
U_{ab}^{(-)}(x,t;y,t'|\,\eta)&=\sum_{q=1}^Lu^*(a,q|\eta,x,t)u(b,q|\eta,y,t')\, , \nonumber \\
U_{ab}^{(+)}(x,t;y,t'|\,\eta)&=\sum_{k=1}^L u(k,b|\eta,x,t)u^*(k,a|\eta,y,t')\, ,\nonumber\\
R_{ab}^{(+)}(x,t;y,t'|\, \eta)&=\overline{e}_a(x,t;y,t'|\, \eta)e_b(x,t;y,t'|\, \eta)\,\nonumber , \\
R_{ab}^{(-)}(x,t;y,t')&=\phi^*_a(x,t)\phi_b(y,t')\, ,
\end{align}
with the auxiliary vectors defined by
\begin{align}
e_a(x,t;y,t'|\, \eta)&=\sum_{k=1}^L u(k,a| \eta,x,t)\phi^*_k(y,t')\, \nonumber  ,\\
\overline{e}_a(x,t;y,t'|\, \eta)&=\sum_{k=1}^Lu^*(k,a| \eta,y,t')\phi_k(x,t).
\end{align}

These determinant expressions capture the exact dynamics 
and provide a  benchmark for validating 
numerical methods. 
The results presented in Fig.~\ref{fig:A_k_w_exact} show details of the single-particle spectral 
properties across different filling fractions. 
For all fillings, the spectral function closely resembles the non-interacting case, 
featuring a well-defined cosine-like dispersion and sharp spectral features. 
The momentum-integrated spectra further illustrate this trend, and display square 
root like singularities at the band-edges.

\bibliography{references}

\end{document}